\documentclass[twocolumn,showpacs,preprintnumbers,amsmath,amssymb,pre]{revtex4}

\usepackage{graphicx}
\usepackage{dcolumn}
\usepackage{bm}
\usepackage{psfrag}

\begin{document}

\preprint{APS/123-QED}

\title{Heat transport in rotating convection without Ekman layers}

\author{S. Schmitz}

\author{A. Tilgner}

\affiliation{Institute of Geophysics, University of G\"ottingen,
Friedrich-Hund-Platz 1, 37077 G\"ottingen, Germany }

\date{\today}

\begin{abstract}
Numerical simulation of rotating convection in plane layers 
with free slip boundaries show that the
convective flows can be classified according to a quantity constructed from the
Reynolds, Prandtl and Ekman numbers. Three different flow regimes appear:
Laminar flow close to the onset of convection, turbulent flow in which the heat
flow approaches the heat flow of non-rotating convection, and an intermediate
regime in which the heat flow scales according to a power law
independent of thermal diffusivity and kinematic viscosity.
\end{abstract}

\maketitle

It is a central problem for many areas of geo- and astrophysics to determine the
heat flux through a rotating and convecting fluid layer. For example, the heat flux through
the atmosphere governs weather and climate, the heat flux through stellar
atmospheres determines stellar evolution, and the heat flux through planetary
cores is essential for the generation of the magnetic fields of these bodies. A
correspondingly large effort has already been spent on the problem. Buoyancy is
driving the flow and can be balanced by either viscous or Coriolis forces, or
the nonlinear terms in the equations of motion, or any combination of these. 
If the Coriolis force dominates the dynamics, a special type of
boundary layer appears near solid boundaries, the Ekman layers, in which the
viscous force is balanced by the Coriolis term. In addition, the flow in the
bulk is organized into columnar vortices with their axes aligned with the rotation axis.
If on the contrary nonlinear advection
supersedes the Coriolis term, these columns are broken up and the style of flow
known from non-rotating convection is approached \cite{Julien96, Stellm04,
King09}. There is an ongoing debate concerning the parameters at which the
transition between these two flow regimes occurs \cite{Canuto98, Ecke99, King09}, and we are still lacking
reliable relations between the heat flux and the control parameters of the flow
that would allow us to extrapolate data from laboratory experiments and
numerical simulation to astrophysical objects.

Some recent work on rotating convection
has focused on the Ekman layers. For instance, ref. \cite{King09} relates
the Ekman layers to the transition mentioned above. Despite the
inhibiting effect of rotation on turbulence, the heat flux in a rotating flow
can exceed that of a non-rotating flow at equal Rayleigh number \cite{Zhong93}. 
In ref. \cite{Zhong09}, this phenomenon is attributed to so called Ekman pumps,
a term reserved for a certain flow pattern associated with Ekman boundary layers
\cite{Cushma94}.
Here we investigate convection with free slip boundary conditions. This
eliminates Ekman layers and one can discern which effect really depends on their
presence. Free slip boundaries are realized to a good approximation in Nature, for example
at the surface of the oceans or at the top of atmospheric layers.

Consider a plane layer of thickness $d$ in the $z-$direction and of infinite
extent in the $x,y-$plane. Let the layer be filled with fluid of kinematic
viscosity $\nu$, thermal diffusivity $\kappa$, and thermal expansion coefficient
$\alpha$. Gravitational acceleration $g$ is pointing in the negative
$z-$direction and the layer is rotating with angular velocity $\Omega$ about the
$z-$axis. The temperatures of the top and bottom boundaries are fixed at $T_0$
and $T_0 + \Delta T$, respectively. These two boundaries are assumed to be free
slip, whereas periodic boundary conditions are applied in the $x-$ and
$y-$directions. The equations of evolution are made non-dimensional by using
$d^2/\kappa$, $d$ and $\Delta T$ for units of time, length, and temperature,
respectively. These equations then become within the Boussinesq approximation
for the dimensionless velocity $\bm v(\bm r, t)$ and temperature $T(\bm r, t)$:

\begin{equation}
\partial_t \bm v + (\bm v \cdot \nabla)\bm v +2 \frac{Pr}{Ek} \hat{\bm z} \times
\bm v = -\nabla p + Pr~ \nabla^2 \bm v
+ Ra~ Pr~ T \hat{\bm z}
\label{eq:NS}
\end{equation}

\begin{equation}
\nabla \cdot \bm v =0
\label{eq:div}
\end{equation}

\begin{equation}
\partial_t T + \bm v \cdot \nabla T 
= \nabla^2 T
\label{eq:temp}
\end{equation}
%
$\hat{\bm z}$ is the unit vector in $z$-direction and $p$ collects the pressure
and the centrifugal acceleration. The boundary conditions require that
$T(z=0)=1$, $T(z=1)=0$, and that $v_z=\partial_z v_x=\partial_z v_y=0$ at both
$z=0$ and $z=1$.
Three independent dimensionless control parameters appear: The Rayleigh number
$Ra$, the Ekman number $Ek$, and the Prandtl number $Pr$. They are defined by:
\begin{equation}
Ra = \frac{g \alpha \Delta T d^3}{\kappa \nu} ~~~,~~~
Ek = \frac{\nu}{\Omega d^2} ~~~,~~~
Pr = \frac{\nu}{\kappa}
\label{eq:RaEkPr}
\end{equation}
The Reynolds number $Re$ and the Nusselt number $Nu$ are an output of the
simulations:
\begin{equation}
Re = \frac{1}{Pr} \sqrt{\frac{1}{V} \int <\bm v^2> dV} ~~~,~~~
Nu = - \frac{1}{A} \int <\partial_z T> dA
\label{eq:ReNu}
\end{equation}
The angular brackets denote average over time and the integrals extend over the
computational volume $V$ for $Re$ and over the surface $A$ of either the top or
the bottom boundary for $Nu$.

The equations of motion were solved with the same spectral method as used in
\cite{Hartle03}, except that free slip boundaries were implemented and
that the Coriolis term was added and treated implicitly
together with the diffusion terms. Resolutions reached up to 129
Chebychev polynomials for the discretization of the $z-$coordinate
and $256 \times 256$ Fourier modes in the $(x,y)-$plane.
The periodicity lengths along the $x-$ and $y-$ directions were always chosen to
be identical.  The aspect ratio, defined as the ratio of the periodicity length
in the $(x,y)-$plane and the layer height, was fixed at 10 for simulations
without rotation. In rotating convection, the typical size of flow structures
varies considerably as a function of the control
parameters, so that it is not useful to use a single aspect ratio. Instead, the
aspect ratio was adjusted for each $Ek$ to fit at least 8 columnar vortices along both the
$x-$ and $y-$directions at the onset of convection, and kept constant as $Pr$ and $Ra$ were varied.

\begin{figure}
\includegraphics[width=8cm]{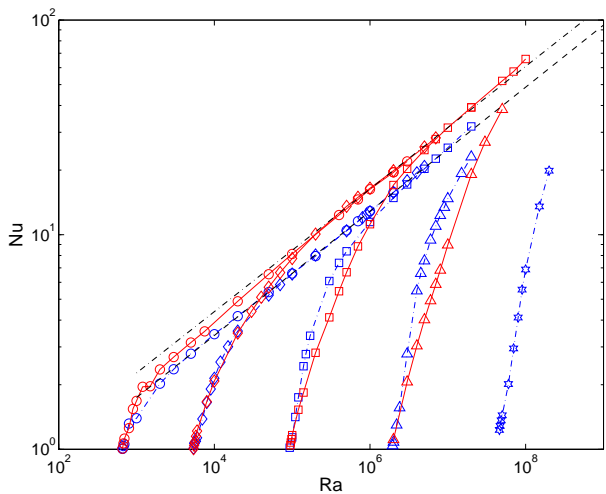}
\caption{$Nu$ as a function of $Ra$ for $Pr=7$ (red symbols and continuous line)
and $0.7$ (blue symbols and dot dashed line), and $Ek=2 \times 10^{-2}$ (diamonds),
$2 \times 10^{-3}$ (squares), $2 \times 10^{-4}$ (triangles) and $2 \times 10^{-5}$ (stars).
Zero rotation
is indicated by circles and the power law fits have an exponent of $0.287$.}
\label{fig1}
\end{figure}

Fig. \ref{fig1} shows $Nu$ as a function of $Ra$ for various $Ek$ and two
different $Pr$. The case of zero rotation is included for comparison. The basic
features visible in this figure are known from previous experiments and
simulations \cite{Rossby69, Zhong93, Julien96, King09}. The onset of convection
is delayed by rotation. After onset, $Nu$ rises more steeply as a function of $Ra$ than in the
non-rotating case. $Nu$ does not follow any simple power in this range of
$Ra$. For large enough $Ra$, the $Nu$ dependence asymptotes towards the
dependence valid for zero rotation, which is well approximated by a power law
in the investigated range of $Ra$.

\begin{figure}
\includegraphics[width=8cm]{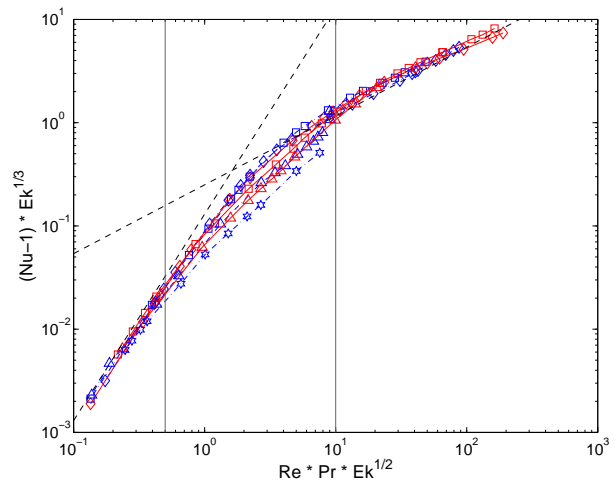}
\caption{$(Nu-1) Ek^{1/3}$ as a function of $Re \, Pr \, Ek^{1/2}$ for the same data
and with
the same symbols as in fig. \ref{fig1}. The dashed lines are power laws with
exponents 2 and 2/3. The two vertical lines indicate the interval outside of
which the fit to one of the two power laws is considered satisfactory.}
\label{fig2}
\end{figure}

All the different curves in fig. \ref{fig1} collapse to a single curve in most of
the parameter range when $(Nu-1) Ek^{1/3}$ is plotted as a function of 
$Re \, Pr \, Ek^{1/2}$ as shown in fig. \ref{fig2}. For large values of $Re \, Pr \, Ek^{1/2}$
one finds $(Nu-1) Ek^{1/3} \propto (Re \, Pr \, Ek^{1/2})^{2/3}$ or $(Nu-1) \propto
(Re \, Pr \,)^{2/3}$. This law is independent of $Ek$ as it should be: At any fixed
$Ek$ and $Pr$, the limit of large $Re$ corresponds to the situation in which the
nonlinear term dominates the Coriolis term, so that one has to recover the
behavior of non-rotating convection, which is of course independent of $Ek$. The
data for zero rotation cannot be included in fig. \ref{fig2} because $Ek$ has no
finite value in this case, but $(Nu-1) \propto (Re \, Pr \,)^{2/3}$ is also found for
strictly zero rotation.

Low values of $Re \, Pr \, Ek^{1/2}$ on the other hand correspond to laminar flows
near the onset of convection. Forming the dot product of eq. (\ref{eq:NS}) and
$\bm v$, integrating over the whole volume and averaging over time, one finds
\begin{equation}
\epsilon = (Nu-1)Ra
\label{eq:epsilon}
\end{equation}
where $\epsilon=\frac{1}{V} \int <(\partial_i v_j) (\partial_i v_j)> dV$ is the adimensional
average dissipation rate of kinetic energy. In a laminar flow, one expects
$\epsilon \propto (Re\, Pr)^2/\lambda^2$, where $\lambda$ is a characteristic
length scale of the flow. For $Pr > 0.676$, convection starts at a critical
Rayleigh number $Ra_c$ obeying $Ra_c \propto Ek^{-4/3}$ and forms stationary cells of size
$\lambda_c$ with $\lambda_c \propto Ek^{1/3}$ \cite{Chandr61}. Eq.
(\ref{eq:epsilon}) becomes $Ek^{-2/3}Re^2 Pr^2 \propto (Nu-1) Ra$. Close to
onset, $Ra \approx Ra_c$ and therefore $(Nu-1) \propto Re^2 Pr^2 Ek^{2/3}$. This
corresponds to the left asymptote in fig. \ref{fig2}. Both the left asymptote
and $(Nu-1) \propto (Re \, Pr \,)^{2/3}$ become straight lines in a logarithmic plot of
$(Nu-1) Ek^{1/3}$ vs. $Re \, Pr \, Ek^{1/2}$, which explains the simple appearance
of fig. \ref{fig2}.

Fig. \ref{fig2} in summary identifies three regimes of rotating convection.
Rotating laminar flow characterizes one of them, and heat transport behaves the
same as in non-rotating convection in another. The transition occurs where the
two asymptotes in fig. \ref{fig2} cross, i.e. at $Re \, Pr \, Ek^{1/2} = 2$. There
is a transition interval around this point of about one decade in width in which
$Nu$ is close to neither asymptote. This third regime will receive detailed
attention below.

\begin{figure}
\includegraphics[width=8cm]{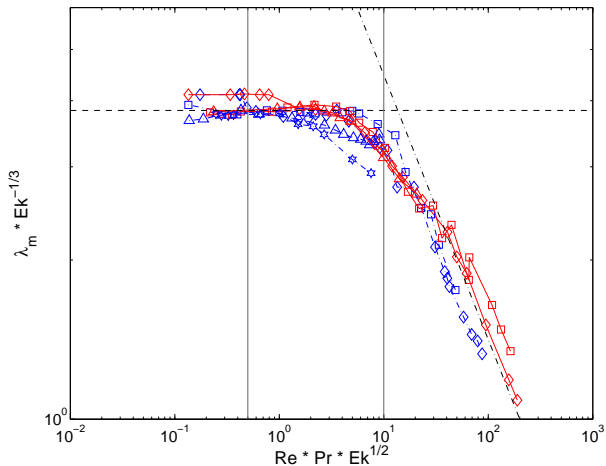}
\caption{$\lambda_m Ek^{-1/3}$ as a function of $Re \, Pr \, Ek^{1/2}$. The symbols
have the same meaning as in fig. \ref{fig1}. The dashed lines indicate power laws with
exponents 0 and -1/2.}
\label{fig3}
\end{figure}

Even though $Nu$ behaves as if there was no rotation for $Re \, Pr \, Ek^{1/2}>10$
in fig. \ref{fig2},
visualizations of the flow still reveal differences. In the rotating case, the
flow forms columnar vortices 
extending from one boundary to the other, whereas for zero rotation, plumes
advected by a large scale circulation are observed. Enough visualizations of
vortices in rotating convection have already appeared \cite{Julien96, Stellm04,
King09} so that there is no need to reproduce any here. The size of the vortices
can be quantified by the method already used in \cite{Hartle03}: Compute the
time averaged advective heat transport through the plane $z=0.5$, $<v_z
\Theta>$, with $\Theta=T-1/2$, compute the Fourier transform of $<v_z \Theta>$, and plot the
spectrum of $<v_z \Theta>$ as a function of wavelength $\lambda$ (see
\cite{Hartle03} for detailed formulas). The median wavelength $\lambda_m$ is
extracted from the spectra, such that the heat advected at wavelengths smaller
than $\lambda_m$ equals the heat advected at larger wavelengths. The value of
$\lambda_m/2$ matches the diameter of the columnar vortices identified visually
in the flow field. Fig. \ref{fig3} shows $\lambda_m Ek^{-1/3}$ as a function of
$Re \, Pr \, Ek^{1/2}$. It is seen that $\lambda_m$ stays at the onset wavelength $\lambda_c$
well into the transition interval and decreases at high $Re \, Pr
\, Ek^{1/2}$. This decrease follows $\lambda_m \propto (Re \, Pr)^{-1/2}$ at
fixed $Ek$, which is compatible with experimental data in \cite{Vorobi98}.

Near the onset of convection, the heat transport is determined by a balance
between buoyancy, Coriolis and diffusive terms. For high $Re \, Pr \, Ek^{1/2}$, the
Coriolis term is overwhelmed by the nonlinear term in eq. (\ref{eq:NS}) so that
$Nu$ is the same as in turbulent, non-rotating convection. Diffusive processes
play a role because all heat has to cross the thermal boundary layers
diffusively. Let us assume as a working hypothesis that the heat flow in the
intermediate regime is governed by a competition between the nonlinear and
Coriolis terms, and that the constraints imposed by rotation on the flow
structure control the heat flux, not diffusion in the boundary layers. The
dimensional heat flow $Q$ must then be given by an expression independent of
$\kappa$ and $\nu$. In order to check this hypothesis, it is convenient to use a
control parameter independent of $\kappa$ and $\nu$. The only combination of
$Ra$, $Ek$ and $Pr$ meeting this requirement is $Ra_*=Ra \, Ek^2/Pr = g \alpha
\Delta T / (\Omega^2 d)$. An appropriate measure of heat flux independent of
$\kappa$ and $\nu$ is $Nu_*= Nu \, Ek/Pr = Q/(\rho c_p \Delta T \Omega d)$, in
which $\rho$ stands for the density and $c_p$ for the heat capacity.

It is useful to replace $Ra_*$ by the flux Rayleigh number $Ra_{f*}$ given by
$Ra_{f*}=Ra_* Nu_* = (g \alpha Q) / (\rho c_p \Omega^3 d^2)$. This combination
is strictly speaking a control parameter only when Neumann conditions are
imposed on the temperature field, which was not the case in our simulations.
However, a parameter based on $Q$ instead of $\Delta T$ is preferable in
astrophysical applications because heat fluxes are better constrained by
observations than vertical temperature differences. We will therefore seek a
relation between $Nu_*$ and $Ra_{f*}$. Furthermore, in a flow dominated by rotation, which
is necessarily nearly two dimensional, it seems plausible that heat flow through a plane
$z=const.$ should be determined solely by the dynamics in that plane. $Q$ would
then be independent of the layer height $d$. If our working hypothesis is
correct that $Q$ is independent of $\kappa$ and $\nu$, and assuming $Nu_*$ is
given by a power law, one has to find a scaling of the form $Nu_* \propto
Ra_{f*}^\beta$. If in addition $Q$ is independent of $d$, one has to find
$\beta=1/2$.

\begin{figure}
\includegraphics[width=8cm]{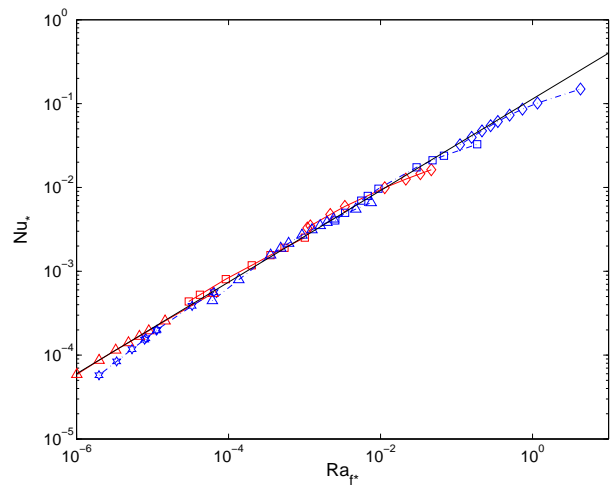}
\caption{$Nu_*$ as a function of $Ra_{f*}$. The symbols have the same meaning as in
fig. \ref{fig1}. This figure contains only those data points which lie in the
interval marked by vertical lines in fig. \ref{fig2}.}
\label{fig4}
\end{figure}

Fig. \ref{fig4} shows $Nu_*$ as a function of $Ra_{f*}$. The figure contains
only those points for which $0.5 < Re\, Pr\, Ek^{1/2} < 10$. This transition
interval is small and does not corroborate any power law $Nu_* \propto
Ra_{f*}^\beta$ at fixed $Ek$ and $Pr$. However, the data for different $Ek$ and
$Pr$ collectively define an envelope which we regard to be the genuine scaling
obeyed by the Nusselt number in the transition regime. The best fit to the data
in fig. \ref{fig4} yields 
\begin{equation}
Nu_*=0.11 \cdot Ra_{f*}^{0.55}
\label{eq:nu*_ra*}
\end{equation}
The exponent $\beta=0.55 \pm 0.01$ is measurably different from $1/2$. There is some scatter
in the points in fig. \ref{fig4} around the power law (\ref{eq:nu*_ra*}). This scatter 
can be reduced by retaining data from a smaller interval of $Re \, Pr \, Ek^{1/2}$, so that
the data are less affected by scalings valid in the neighboring intervals.

Ref. \cite{Christ02} investigates thermal convection in a rotating spherical
shell. In this geometry, convection occurs mostly outside a cylinder tangent to
the inner core and coaxial with the rotation axis, whereas the flow velocities
are much smaller inside the tangent cylinder. Gravitational acceleration varies
radially in the simulations in ref. \cite{Christ02} and there is a zonal flow along
circles of constant latitude which has no analog in our simulations. Despite
all these differences, the heat flux in the spherical geometry obeys
$Nu_*=0.077 \cdot Ra_{f*}^{5/9}$ according to ref. \cite{Christ02} and
the best fit to a compilation of data 
in ref. \cite{Aurnou07} yields $Nu_*=0.08 \cdot Ra_{f*}^{0.55}$.
The exponent in (\ref{eq:nu*_ra*}) appears to be very robust.

It is also interesting to draw a parallel with dimensional arguments for
non-rotating convection \cite{Spiege71}. If the heat transfer is independent of the layer
thickness because it is determined by
boundary layer dynamics, $Nu$ has to behave like $Nu \propto Ra^{1/3}$. This exponent is
generally not observed experimentally because of the presence of a
large scale circulation. The assumption that heat transport is independent of
thermal diffusivity and kinematic viscosity leads without rotation to $Nu
\propto (Ra \, Pr \,)^{1/2}$. While this scaling has been found in simulations
avoiding boundary layers \cite{Lohse03} it remains elusive in any bounded
geometry. In rotating convection, fig. \ref{fig4} shows that a power law independent of
diffusivities is a useful fit to the data, but the heat flow still depends on the layer depth.

In summary, three different regimes of convection could be identified as a
function of $Re \, Pr \, Ek^{1/2}$. For small and large values of 
$Re \, Pr \, Ek^{1/2}$, one approaches asymptotically the scalings valid for
rotating convection near onset and non-rotating convection, respectively. The
cross-over occurs in a transition interval around $Re \, Pr \, Ek^{1/2}=2$. 
This contradicts the naive expectation that
the transition should occur when the Rossby number $Ro=Re \, Ek$ equals 1. Even
though $Re$ is not a control parameter, the transition criterion is useful when
observations yield some information about the flow velocities in a celestial body. A case in
point is the Earth's core, for which magnetic secular variations provide us with
estimates of typical flow velocities around $5 \times 10^{-4} m/s$. Together with
$\Omega=7.29 \times 10^{-5} s^{-1}$ and the generally accepted material
properties inside the core of $\kappa=3 \times 10^{-6} m^2/s$ and $\nu= 5 \times
10^{-7} m^2/s$ \cite{Schube07}, one finds $Re \, Pr \, Ek^{1/2} \approx 5$,
which places the Earth's core inside the transition interval. If on the other hand 
the Earth's core is driven by compositional convection, a diffusivity of $7
\times 10^{-9} m^2/s$ should be used \cite{Schube07},
leading to $Re \, Pr \, Ek^{1/2} \approx 6 \times 10^3$.

\acknowledgments
This work was supported by the Deutsche Forschungsgemeinschaft (DFG).


\end{document}